\documentclass[conference]{IEEEtran}
\usepackage{fancyhdr}
\IEEEoverridecommandlockouts

\usepackage{cite}
\usepackage{amsmath,amssymb,amsfonts,booktabs}
\usepackage{graphicx}
\usepackage{textcomp}
\usepackage{xcolor}
\usepackage{xspace}
\usepackage{subcaption}
\usepackage{multirow}
\usepackage{enumitem}
\usepackage{tabularx} 
\usepackage{listings}
\usepackage{algorithm}
\usepackage{algorithmicx}
\usepackage{xfrac}
\usepackage{hyperref}
\usepackage{algpseudocode}
\usepackage{float}
\usepackage{tikz}
\def\BibTeX{{\rm B\kern-.05em{\sc i\kern-.025em b}\kern-.08em
    T\kern-.1667em\lower.7ex\hbox{E}\kern-.125emX}}

\newcommand{\spsyrk}{\textsc{SpSYRK}\xspace}
\newcommand{\commspsyrk}{\textsc{CommSpSYRK}\xspace}

\definecolor{revblue}{RGB}{0,0,255}
\newcommand{\rev}[1]{{#1}}
\renewcommand{\rev}[1]{\textcolor{black}{#1}}

\newcommand{\TM}[1] {{#1}}

\newcommand{\B}{\mathbf{B}\xspace}
\newcommand{\A}{\mathbf{A}\xspace}
\newcommand{\C}{\mathbf{C}\xspace}
\newcommand{\R}{\mathbb{R}}
\newcommand{\cO}{\mathcal{O}}
\newcommand{\T}{\mathsf{T}\xspace}

\definecolor{darkgreen}{RGB}{0,100,0}
\newcommand{\new}[1]{{#1}}
\definecolor{darkorange}{rgb}{0.75,0.35,0.0}

\begin{document}

\title{SpSYRK: Half the Work in Distributed Sparse Matrix Multiplication} 

\author{
\IEEEauthorblockN{Thomas McFarland}
\IEEEauthorblockA{\textit{Cornell University}\\
Ithaca, NY, USA \\
tfm62@cornell.edu}
\and
\IEEEauthorblockN{Julian Bellavita}
\IEEEauthorblockA{
\textit{Cornell University}\\
Ithaca, NY, USA \\
jb2695@cornell.edu}
\and
\IEEEauthorblockN{Giulia Guidi}
\IEEEauthorblockA{
\textit{Cornell University}\\
Ithaca, NY, USA \\
gg434@cornell.edu}
}


\maketitle
\thispagestyle{fancy}
\lhead{}
\rhead{}
\chead{}
\lfoot{\footnotesize{
SC26 Workshops, November 15-20, 2026, Chicago, Illinois, USA
\newline 979-8-3195-1221-5/26/\$31.00 \copyright 2026 IEEE}}
\rfoot{}
\cfoot{}
\renewcommand{\headrulewidth}{0pt}
\renewcommand{\footrulewidth}{0pt}

\begin{abstract}
The symmetric rank-$k$ update (SYRK), $\C = \A \A^\top$, computes the dot product of each pair of rows of $\A$, producing the Gram matrix $\C$.
Its sparse variant underpins similarity search in machine learning, graph analytics, and genomics, including Jaccard similarity on datasets too large to fit on a single machine node.
\rev{Despite the symmetry in its inputs and outputs, existing distributed sparse matrix multiplication algorithms, such as Sparse SUMMA, treat sparse SYRK as generic multiplication: they compute the full output and materialize the explicit transpose, even when the calling application uses only one triangle. 
Prior distributed $\A\A^\top$ computations in similarity search and genome assembly inherit this overhead from the underlying SpGEMM.}

In this paper, we present \textsc{SpSYRK} and \textsc{CommSpSYRK}, two distributed sparse SYRK algorithms that exploit symmetry.
The first, \textsc{SpSYRK}, partitions the off-diagonal blocks of the output between the upper and lower triangular regions of the process grid and computes only the lower-triangular part of each diagonal block, halving per-process computation compared with state-of-the-art distributed SpGEMM.
The second, \textsc{CommSpSYRK}, further reorders communication to avoid forming $\A^\top$, which reduces per-process communication volume.

On 32 nodes of the Perlmutter supercomputer, \spsyrk achieves a 2$\times$ speedup over an optimized Sparse SUMMA on matrices where local multiplication dominates the runtime; the advantage narrows on communication-bound inputs, a dependence that the cost model predicts from the arithmetic intensity.
\textsc{CommSpSYRK} fixes this and consistently achieves superior scaling at high process counts.
The approach is a drop-in replacement for any application computing $\C = \A \A^\top$ via a distributed SpGEMM routine, and its triangular output can be consumed directly by subsequent operations, reducing both computation and memory footprint.
\end{abstract}





\section{Introduction}

Computing the product of a large sparse matrix with its transpose, $\C = \A\A^\top$ for $\A \in \R^{m \times n}$, is a core operation in many scientific and engineering applications.
This operation is known as the sparse symmetric rank‑$k$ update, or sparse SYRK.
It appears in computing Jaccard similarity~\cite{besta2020communication}, hypergraph partitioning~\cite{hussain_communication-avoiding_2021}, triangle counting~\cite{azad2015parallel}, the normal equations $\A^\top\A$ in least-squares problems, Gram matrix construction in statistics and machine learning, genome assembly~\cite{guidi2021bella, guidi2021parallel}, and protein sequence alignment and similarity search~\cite{buluc2025ubiquitoussparsematrixmatrixproducts, selvitopi2020distributed, selvitopi2022extreme}.
In practice, $\A$ often has hundreds of millions of nonzeros, and the output $\C$ can have up to two orders of magnitude more nonzeros than $\A$ (Table~\ref{tab:sizes}).
In addition, the number of FLOPS required to compute $\C$ is at least as large as the number of nonzeros in $\C$, so significant computation is also needed. 
In these settings, distributed-memory parallelism is required, and applications typically use a distributed-memory sparse general matrix-matrix multiply (SpGEMM) routine to perform sparse SYRK.

This strategy fails to exploit the fact that $\C$ is symmetric. 
In particular, off-the-shelf distributed SpGEMM routines, such as CombBLAS~\cite{azad2021combinatorial}, compute the entire output matrix $\C$, even though only the lower or upper triangular portion of $\C$ is actually needed.
They also materialize the explicit transpose $\A^\top$, incurring additional communication and memory costs that, as we show, are unnecessary.

Here, we introduce and implement \spsyrk, a distributed-memory sparse SYRK algorithm that exploits output symmetry in the multiplication to avoid redundant computation.
Distributed sparse $\A\A^\top$ has been computed before, notably in the Jaccard, protein-similarity, and genome-assembly pipelines cited above, but those approaches use a general distributed SpGEMM; that is, they form the full symmetric output and explicitly materialize $\A^\top$.
Our key novelty is a SYRK-aware multiplication that never computes the redundant triangular part and, in the \textsc{CommSpSYRK} variant, never materializes $\A^\top$.
\spsyrk uses a 2D partitioning scheme similar to the Sparse SUMMA algorithm~\cite{buluc_combinatorial_2011} for distributed SpGEMM, but exploits output symmetry by splitting the computation of off-diagonal blocks between processes in the upper and lower triangular regions of the process grid, while diagonal processes compute only the lower triangular portion of their block, theoretically halving the computation.

A main component is \textsc{CommSpSYRK}, a communication-reordered variant that eliminates the explicit materialization of $\A^\top$.
Rather than transposing and redistributing the input, \textsc{CommSpSYRK} reorders the communication so that each process receives only the submatrices it needs.
This variant has the lowest communication volume of the schemes presented and delivers the best performance at scale.
The regions of the input matrices that must be communicated to each process are smaller, reducing asymptotic per-process communication volume to $\sfrac{3}{4}$ of that of Sparse SUMMA.
The algorithm can also materialize the full output matrix from the explicitly computed half, which is necessary for downstream tasks such as the overlap detection component of ELBA~\cite{guidi2021parallel}.
This materialization is often wasteful, though; many applications work directly with the triangular output, which always offers a memory advantage and, in specific pipelines, a computational one.

The evaluation compares strong scaling against the distributed SpGEMM routine in CombBLAS~\cite{azad2021combinatorial} and an optimized Sparse SUMMA implementation that is \TM{usually} faster than the CombBLAS version. 
Both were included to provide an independent algorithm and a SUMMA algorithm using the same kernels.
On 32 nodes of NERSC Perlmutter, \textsc{SpSYRK} achieves twice the performance of optimized Sparse SUMMA on compute-bound inputs, and both \textsc{SpSYRK} and \textsc{CommSpSYRK} are up to $10\times$ faster than CombBLAS.
On bandwidth-bound inputs, relative to optimized Sparse SUMMA, \commspsyrk achieves up to a $2\times$ speedup.

\rev{The main contributions are:
\begin{enumerate}
    \item \textsc{SpSYRK}, a distributed-memory SYRK algorithm that exploits output symmetry in the multiplication to avoid the redundant triangle;
    \item \textsc{CommSpSYRK}, a communication-reordered variant that eliminates the explicit materialization of $\A^\top$ and achieves the best performance at scale;
    \item A triangular output format that provides memory advantages for symmetric outputs and computational benefits in specific pipelines;
    \item An evaluation on 32 Perlmutter nodes showing that \commspsyrk is up to $2\times$ faster than Sparse SUMMA and up to $10\times$ faster than CombBLAS.
\end{enumerate}}
The repository can be found at \url{https://github.com/CornellHPC/SpSYRK}.

\section{Background and Related Work}

In this section, we define sparse SYRK, review the SUMMA algorithm and prior work on distributed SpGEMM.

\subsection{SpGEMM and Sparse SYRK}

SpGEMM computes $\C = \A\B$ where both inputs are sparse, and is a widely used primitive~\cite{10.1007/978-3-540-75755-9_32, 8547538, besta2020communication, guidi2021bella, selvitopi2020distributed, bellavita2026communication, li2026ocean}.
This paper focuses on the sparse symmetric rank-$k$ update, or sparse SYRK: given a sparse matrix $\A \in \R^{m \times n}$, compute $\C = \A\A^\top \in \R^{m \times m}$.
The output $\C$ is symmetric, so only its lower (or upper) triangular part contains unique information.
The sparse SYRK is a special case of SpGEMM in which the second operand is the transpose of the first.

In the dense setting, exploiting symmetry in rank-$k$ updates is standard.
SYRK is a level-3 BLAS routine~\cite{blas3, dongarra1990set}, and its distributed counterpart \texttt{PxSYRK} in PBLAS, used throughout ScaLAPACK~\cite{blackford1997scalapack}, computes only one triangle and halves the arithmetic relative to a general matrix multiplication.
\rev{There is no analogous routine in distributed sparse memory: sparse libraries expose only general SpGEMM, so applications that compute $\A\A^\top$ form the full output and discard half.
This is the case for earlier distributed SpGEMM applications that compute a symmetric product, such as the similarity search and genome assembly pipelines above, where the redundant triangle and explicit transpose are inherited from the general SpGEMM they invoke.}
This work closes that gap by addressing challenges absent in the dense case, such as unpredictable output sparsity and load imbalance caused by irregular nonzero distribution.

\begin{algorithm}[t]
\begin{algorithmic}[1]
\Require $\mathbf{A}$ and $\mathbf{B}$ distributed on a $p_r \times p_c$ process grid
\Ensure Matrix $\mathbf{C} = \mathbf{A}\mathbf{B}$
\State $(i,j) \gets$ grid indices of this process
\State Initialize $\mathbf{C}_{\text{local}} = 0$
\For{$\text{stage} = 0 \to p_c - 1$}
    \If{$j = \text{stage}$}
        \State Broadcast $\mathbf{A}_{\text{local}}$ within process row
    \Else
        \State Receive $\mathbf{A}_{\text{recv}}$ from process $(i, \text{stage})$
    \EndIf
    \If{$i = \text{stage}$}
        \State Broadcast $\mathbf{B}_{\text{local}}$ within process column
    \Else
        \State Receive $\mathbf{B}_{\text{recv}}$ from process $(\text{stage}, j)$
    \EndIf
    \State $\mathbf{C}_{\text{local}} \mathrel{+}= \mathbf{A}_{\text{recv}} \new{\,} \mathbf{B}_{\text{recv}}$
\EndFor
\State \Return $\mathbf{C}_{\text{local}}$
\end{algorithmic}
\caption{Sparse SUMMA from the view of process $(i, j)$}
\label{alg:summa}
\end{algorithm}

\subsection{Distributed SpGEMM}

Prior work has developed several distributed-memory SpGEMM algorithms.
Sparse SUMMA~\cite{buluc_combinatorial_2011} and Sparse Cannon~\cite{sparse-cannon} adapt the standard SUMMA and Cannon algorithms to sparse inputs, partitioning the inputs across a 2D process grid and casting SpGEMM as a sequence of block outer products.
Both have scalable communication costs and rely on random permutation for load balancing, a strategy also used here. 
Sparse SUMMA is the basis for the SpGEMM implementation in CombBLAS~\cite{azad2021combinatorial}.

Brock et al.~\cite{brock2024rdma} introduce a GPU-accelerated Sparse SUMMA that uses one-sided communication and work stealing for load balancing without random permutation.
Bellavita et al.~\cite{bellavita2026communication} present a distributed SpGEMM for GPU machines with hierarchical interconnects, using a hybrid 1D/2D trident partitioning and a communication schedule that reduces off-node traffic.
Both target GPU platforms, which makes direct comparison difficult, and are largely orthogonal to this work.

Hussain et al.~\cite{hussain_communication-avoiding_2021} present a Sparse SUMMA variant for outputs too large for memory, generating a subset of output columns at a time to support applications that never require the full output, such as Markov clustering.
Hong et al.~\cite{hong2024sparsity} use 1D partitioning with sparsity-aware communication via one-sided MPI, scaling better than Sparse SUMMA only when the input is highly structured. 
PETSc and Trilinos implement a similar 1D, point-to-point, sparsity-aware approach~\cite{petsc, trilinos}.
Ranawaka et al.~\cite{ranawaka2024distributed} develop a distributed SpGEMM algorithm for $\C = \A\B$, where $\B$ is tall and skinny.
\rev{Notably, none of these works exploit symmetry in the case of $\C = \A \A^T$ to reduce redundant computation or communication.}

\section{Algorithm}\label{sec:alg}

Our goal is to compute the sparse SYRK operation $\C = \A\A^T$ for a sparse matrix $\A \in \R^{m \times n}$, distributed across $p$ processes organized in a $q \times q$ grid, where $q = \sqrt{p}$. 
For simplicity, we assume a square process grid and square blocks; results for rectangular inputs are discussed in Section~\ref{sec:runtime-rect}.
The square-grid requirement is inherited from the CombBLAS backend~\cite{buluc_combinatorial_2011}; individual processes hold rectangular submatrices when $\A$ is non-square.
Process $(i,j)$ owns an $\sfrac{m}{q} \times \sfrac{m}{q}$ block $\C_{ij}$ of the output and an $\sfrac{m}{q} \times \sfrac{n}{q}$ block $\A_{ij}$ of the input.
Our starting point is Sparse SUMMA (Alg.~\ref{alg:summa}), which computes $\C = \A\B$ with $\B = \A^T$ materialized explicitly: at stage $s$, process $(i,s)$ broadcasts $\A_{is}$ along its process row and process $(s,j)$ broadcasts $\B_{sj}$ along its process column, and each process accumulates $\C_{ij} = \sum_s \A_{is}\B_{sj}$.

If we apply this to sparse SYRK, it is wasteful because (i) $\C$ is symmetric, so computing all of it does twice the necessary work, and (ii) $\B = \A^\top$ contains no information not already in $\A$, yet the baseline materializes it via a transpose (a local operation plus a communication round), which doubles the memory overhead of the input.
\textsc{SpSYRK} eliminates the first redundancy through symmetry-aware partitioning of computation, and \textsc{CommSpSYRK} additionally eliminates the second through a reordered communication schedule that never forms $\A^\top$. 
Both variants are agnostic to the local kernel: any local sparse SYRK or SpGEMM routine can be used, so the algorithm ports across hardware and library ecosystems.

\subsection{Symmetry-Aware Computation}\label{sec:alg-comp}

The symmetry $\C_{ij} = \C_{ji}^\top$ lets each process compute only half of its assigned block. 
For a diagonal process $(i,i)$, the local accumulation is
\[
 \C_{ii} = \sum_s \A_{is}(\A^\top)_{si}
 = \sum_s \A_{is}\A_{is}^\top
 = \sum_s \mathrm{SYRK}(\A_{is}),
\]
a sum of local rank-$k$ updates, each with a symmetric result. 
Diagonal processes therefore compute only the lower (or equivalently, upper) triangle of $\C_{ii}$.

\begin{figure}
    \centering
    \includegraphics[width=\linewidth]{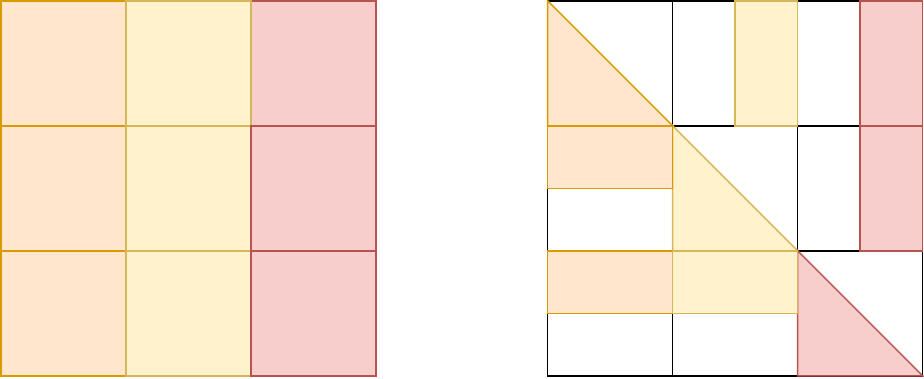}
    \caption{Computation of each process in Sparse SUMMA (left) versus \textsc{CommSpSYRK} (right) on a $3\times 3$ grid. Colors indicate communicators for the broadcast of $\A^T$.
    Diagonal processes compute only the lower triangle of their block; off-diagonal processes compute the complementary halves.
    \new{Please note that for demonstration purposes, the blocks here are dense, whereas in practice they will be sparse with varying numbers of nonzeros.}
    }
    \label{fig:compscheme}
\end{figure}

For an off-diagonal pair, $\C_{ij}$ on process $(i,j)$ is the transpose of $\C_{ji}$ on $(j,i)$, so the work is split evenly.
The row/column split is chosen to reduce computation: the lower-triangular process $(i,j)$ with $i > j$ computes the first $\tfrac{m}{2q}$ rows of $\C_{ij}$, and its partner $(j,i)$ computes the last $\tfrac{m}{2q}$ columns of $\C_{ji}$. 
Given that row $r$ of $\C_{ij}$ is column $r$ of $\C_{ji}$, the two half-blocks are disjoint and together cover the full block pair (Figure~\ref{fig:compscheme}).
In a load-balanced matrix, every process computes half of its Sparse SUMMA workload, with the remainder recovered by symmetry for diagonal processes or from the partner for off-diagonal processes.
A load-balanced computation can be induced via a symmetric random permutation, which our implementation supports.
\rev{The even split refers to the assigned blocks, not the nonzeros within them.
The halves are geometrically equal, but their nonzero counts diverge for sparse inputs, and the divergence increases with problem size. 
The symmetric random permutation bounds this by spreading nonzeros uniformly in expectation; Sections~\ref{sec:load} and~\ref{sec:strongscaling} evaluate its impact.}

On completion, each process holds half of its output block, namely the lower triangle on the diagonal and the complementary half-blocks off the diagonal.
If an application requires the full matrix, \textsc{SpSYRK} materializes it by transposing.
In many applications, however, the triangular output is used directly.
Thus, skipping materialization halves the memory of $\C$, which is typically much larger than $\A$, and removes the output transpose from the performance-critical path.

\subsection{\textsc{CommSpSYRK}: Reduced-Volume Communication}\label{sec:alg-comm}

\begin{figure}
    \centering
    \includegraphics[width=\linewidth]{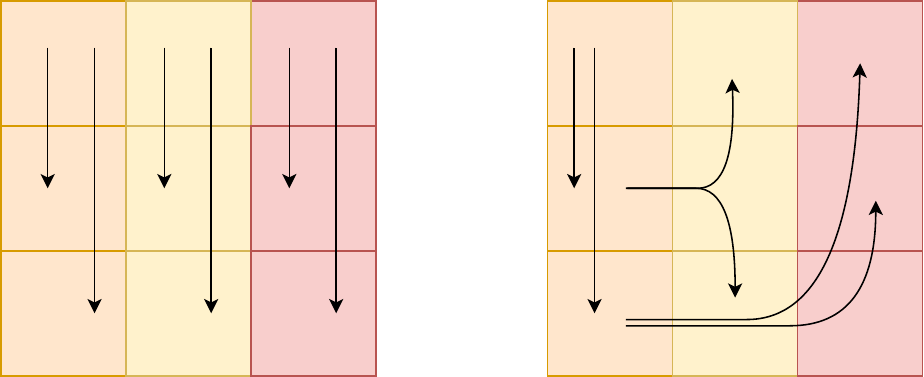}
    \caption{Communication at stage 0 in Sparse SUMMA (left) versus \textsc{CommSpSYRK} (right) on a $3\times 3$ grid. In \textsc{CommSpSYRK}, the column-direction data originates from the transposed grid positions and is trimmed to each receiver's needs.}
    \label{fig:commscheme}
\end{figure}

The computation split reduces the inputs required by each process. Given that the first $\tfrac{m}{2q}$ rows of $\C_{ij} = \sum_s \A_{is} (\A^\top)_{sj}$ depend only on the first $\tfrac{m}{2q}$ rows of each $\A_{is}$, and the last $\tfrac{m}{2q}$ columns of $\C_{ji}$ depend only on the last $\tfrac{m}{2q}$ 
columns of each $(\A^\top)_{sj}$, lower-triangular processes need only the top half of each received $\A$ block, and upper-triangular processes need only the right half of each received $\A^\top$ block.
Diagonal processes need no $\A^\top$ at all, since their accumulation involves only products of the form $\A_{is}\A_{is}^\top$.
One iteration of the resulting pattern is shown on the right in Figure~\ref{fig:commscheme}. 
The overall effect is that less data needs to be communicated at each stage.

Notably, implementing this system cannot be done using MPI collectives.
MPI\_Bcast sends identical data to all receivers, MPI\_Scatterv splits the root’s buffer into disjoint pieces, and MPI\_Alltoallv requires every member of a communicator to participate.
Thus, we use MPI\_ISend and MPI\_IRecv to implement \textsc{CommSpSYRK}.

\commspsyrk also eliminates $\A^\top$.
Observe that in the baseline, the block $(\A^T)_{sj} = \A_{js}^T$ broadcast at stage $s$ is simply the transpose of a block of $\A$ that already resides on process $(j,s)$.
\textsc{CommSpSYRK} reorders the send pattern: at stage $s$, instead of process $(s,j)$ sending $\B_{sj}$ down column $j$, process $(j,s)$ sends $\A_{js}$ down column $j$.
The set of receivers is unchanged; only the sender is reflected across the diagonal.
In other words, the same processes need to have $\B$ communicated to them (including the original sender), but the sender is instead located at the transpose of the original sender’s position.
This is illustrated in Figure~\ref{fig:commscheme}.
The local accumulation becomes:

$$\C_{ij} = \sum_s \A_{is}\A_{js}^T,$$

\noindent
pushing the transpose into the local kernel, where it can be fused with the multiply at negligible cost.
Thus, \textsc{CommSpSYRK} never materializes $\A^\top$.
This rules out collectives, since each receiver gets a different message and the sender is outside any fixed broadcast group, so \commspsyrk uses $q - 1$ point-to-point messages per stage instead of an $\cO(\log q)$ depth broadcast.
\subsection{Theoretical Analysis}\label{sec:analysis}

The analysis uses the $\alpha$–$\beta$ model~\cite{hockneyCommChallenges}, extended with local computation rates.
A message of $w$ words costs $\alpha_{\mathrm{comm}} + \beta_{\mathrm{comm}} w$, where $\alpha_{\mathrm{comm}}$ is the per-message latency and $\beta_{\mathrm{comm}}$ is the per-word transfer time.
\new{A pipelined tree broadcast~\cite{bruck1992multiple} sends $\log(p)$ messages and $\mathcal{O}(n)$ words for a broadcast of size $n$ among $p$ processors.}
Each local kernel $x \in \{\mathrm{mul}, \mathrm{tr}\}$ costs $\alpha_{\mathrm{loc}} + \beta_x u_x$, with $\alpha_{\mathrm{loc}}$ a shared per-invocation overhead and $\beta_x$ the marginal cost per unit of work $u_x$: multiply-adds for the local multiply, nonzeros processed for the sparse addition and transpose.
Let $\eta = \mathrm{nnz}(\A)/p$ be the per-process nonzero count, which we assume is uniform for the sake of analysis; $m' = m/q$ the local row count; $\gamma = \mathrm{nnz}(\C)/\mathrm{nnz}(\A)$ the output expansion factor; and
\[
  \varphi = \frac{1}{p}\sum_{k=1}^{n}\mathrm{nnz}\bigl(\A(:,k)\bigr)^2
\]
\rev{the per-process multiply-add count of $\A\A^\top$ under random permutation.}
$\varphi$ is not proportional to $\eta$, since it depends on the column-nonzero distribution: matrices with equal $\mathrm{nnz}(\A)$ can differ by orders of magnitude in local multiplication cost.

For simplicity, $\A$ is assumed square ($m = n$). 
The three components are:
\begin{align}
  T_{\mathrm{tr}}(\nu)   &= \underbrace{\alpha_{\mathrm{loc}} +
      \beta_{\mathrm{tr}}\nu}_{\text{local transpose}} +
      \underbrace{\alpha_{\mathrm{comm}} +
      \beta_{\mathrm{comm}}(2\nu + m')}_{\text{pairwise exchange}}, \\
  T_{\mathrm{comp}}(f)   &= \alpha_{\mathrm{loc}} + \beta_{\mathrm{mul}}f, \\
  T_{\mathrm{comm}}(\nu) &= c\,\alpha_{\mathrm{comm}} +
      \beta_{\mathrm{comm}}(2\nu + m'), \label{eq:comm}
\end{align}
where a CSR block with $\nu$ nonzeros occupies $2\nu + m'$ words, and $c$ is the message multiplier: $c = 2\log q$ for a broadcast and $c = q-1$ for a point-to-point round.
\rev{Every variant runs exactly $q$ stages; $c$ multiplies only $\alpha_{\mathrm{comm}}$, as in Eq.~\ref{eq:comm}.}

Sparse SUMMA, \textsc{SpSYRK}, and \textsc{CommSpSYRK} share one structure: an optional input transpose, $q$ stages of communicating two blocks and a local multiply-accumulate, and an optional output transpose.
They differ in where and how much data moves:
\begin{multline}\label{eq:cost-model}
  T = \tau_{\mathrm{in}}\,T_{\mathrm{tr}}(\eta)
    + T_{\mathrm{comp}}(\sigma_{\mathrm{mul}}\varphi) \\
    + q\bigl[T_{\mathrm{comm}}(\sigma_{\A}\eta)
    + T_{\mathrm{comm}}(\sigma_{\B}\eta)\bigr]
    + \tau_{\mathrm{out}}\,T_{\mathrm{tr}}(\sigma_{\mathrm{mul}}\gamma\eta).
\end{multline}
The coefficients $\tau_{\mathrm{in}}, \tau_{\mathrm{out}} \in \{0,1\}$ toggle the transposes, $\sigma_{\mathrm{mul}}$ scales local work, and $\sigma_{\A}, \sigma_{\B}$ scale the communicated volume of each operand. 
Tables~\ref{tab:valsglob} and~\ref{tab:valscomm} give these by algorithm, and, for \textsc{CommSpSYRK}, by grid position.

\begin{table}
    \centering
    \renewcommand{\arraystretch}{1.15}
    \begin{tabular}{c||c|c|c}
        & Sparse SUMMA & \spsyrk & \commspsyrk \\
        \hline
        $c$ & $2\log q$ & $2\log q$ & $q-1$ \\
        $\tau_{\mathrm{in}}$ & $1$ & $1$ & $0$ \\
        $\tau_{\mathrm{out}}$ & $0$ & $0$ or $1^{\dagger}$ & $0$ or $1^{\dagger}$ \\
        $\sigma_{\mathrm{mul}}$ & $1$ & $\sfrac{1}{2}$ & $\sfrac{1}{2}$
    \end{tabular}
    \caption{Cost coefficients by algorithm. $^{\dagger}$Output transpose applies only when the application requires full materialization of $\C$; the triangular-output mode sets $\tau_{\mathrm{out}} = 0$.}
    \label{tab:valsglob}
\end{table}

\begin{table}
    \centering
    \renewcommand{\arraystretch}{1.15}
    \begin{tabular}{c||c|c|c}
        & Upper & Lower & Diagonal \\
        \hline
        $\sigma_{\A}$ & $1$ & $\sfrac{1}{2}$ & $1$ \\
        $\sigma_{\B}$ & $\sfrac{1}{2}$ & $1$ & $0$ \\
        Received volume & $\sfrac{3}{4}$ & $\sfrac{3}{4}$ & $\sfrac{1}{2}$
    \end{tabular}
    \caption{Communication scaling factors for \commspsyrk by grid position, relative to Sparse SUMMA.
    For \spsyrk, $\sigma_{\A} = \sigma_{\B} = 1$ everywhere.}
    \label{tab:valscomm}
\end{table}

Because $\sigma_{\mathrm{mul}} = \sfrac{1}{2}$, both \textsc{SpSYRK} and \textsc{CommSpSYRK} cut the local work in half.
\textsc{CommSpSYRK} further reduces communication: off-diagonal processes receive $\sfrac{3}{4}$ of the baseline volume and diagonal processes $\sfrac{1}{2}$, so with diagonal processes occupying only a $\sfrac{1}{q}$ fraction of the grid, the asymptotic per-process volume is $\sfrac{3}{4}$ that of Sparse SUMMA, and the input transpose is eliminated.

\rev{As $q$ grows, local work falls as $1/q^2$, communicated volume as $\mathrm{nnz}(\A)/q$, and message count grows as $2q\log q$ (broadcast) or $q(q-1)$ (point-to-point). 
A small $q$ is compute-bound, where \textsc{CommSpSYRK} also saves the input transpose; intermediate $q$ is bandwidth-bound, where \textsc{CommSpSYRK}'s reduced volume wins and \textsc{SpSYRK} converges to Sparse SUMMA; large $q$ is latency-bound, where \textsc{CommSpSYRK}'s $q-1$ messages become a liability against a $2\log q$ broadcast.}
\rev{This last crossover occurs near $\beta_{\mathrm{comm}}\eta \approx \alpha_{\mathrm{comm}}(q-1-2\log q)$. Calibrated on Perlmutter, where $\alpha_{\mathrm{comm}}\approx 4\times 10^{-6}\,\mathrm{s}$ and $\beta_{\mathrm{comm}}\approx 4\times 10^{-10}\,\mathrm{s/word}$, it lands at $q \approx 32$ for the sparsest input (\texttt{rgg}) and higher elsewhere, beyond $\approx 1024$ processes.}
This is consistent with the results presented in Section~\ref{sec:strongscaling}.
\section{Implementation and Optimization}
\label{sec:impl}

Our implementation relies on CombBLAS for I/O and baselines~\cite{azad2021combinatorial}, MPI for inter-process communication, and Intel MKL for local computation~\cite{wang2014intel}.
\new{Our implementation spans four configurations across two independent axes.
The first axis selects the communication pattern, either the Sparse SUMMA broadcast or the point-to-point \textsc{CommSpSYRK} schedule, isolating the communication savings of the new schedule from the computational savings of the SYRK-aware kernel.
The second axis selects whether $\C$ is fully materialized or stored halved, isolating the memory savings from the triangular output.}

\begin{algorithm}[t]
\caption{Local \spsyrk computation on process $(i,j)$}
\label{alg:localcomp}
\begin{algorithmic}[1]
\Require Left block $\A_{\mathrm{recv}}$; right block $\mathbf{R}_{\mathrm{recv}}$ when $i \neq j$.
  \Statex In \spsyrk, $\mathbf{R}_{\mathrm{recv}}$ is a received block of $\A^\top$;
  \Statex In \commspsyrk, $\mathbf{R}_{\mathrm{recv}} = \A'^{\top}_{\mathrm{recv}}$, transposed in-kernel.
\Ensure Accumulates this process's half of $\C_{ij}$
\If{$i = j$}
    \State $\C_{\mathrm{local}} \mathrel{+}= \textsc{syrk}(\A_{\mathrm{recv}})$ \Comment{lower triangle only}
\ElsIf{$i > j$}
    \State $\C_{\mathrm{local}} \mathrel{+}= \A_{\mathrm{recv}}[\,{:}\,r/2,\ :]\ \times\ \mathbf{R}_{\mathrm{recv}}$ \Comment{Row top half}
\Else
    \State $\C_{\mathrm{local}} \mathrel{+}= \A_{\mathrm{recv}}\ \times\ \mathbf{R}_{\mathrm{recv}}[\,:,\ c/2\,{:}\,]$ \Comment{Col right half}
\EndIf
\end{algorithmic}
\end{algorithm}

\begin{algorithm}
\caption{\textsc{CommSpSYRK} send at stage $s$, process $(i,j)$}
\label{alg:commsched}
\rev{
\begin{algorithmic}[1]
\Require $\A$ distributed on a $q \times q$ grid
\Ensure Each process receives the input blocks required for its local computation
\Statex \textbf{Row send} (supplies $\A_{is}$ along process row $i$):
\If{Process is in row sender stage ($j=s$)}
    \For{each receiver $(i, j')$ in row $i$}
        \If{Receiver is upper-triangle ($i \leq j'$)}
            \State \textsc{ISend} $\A_{\mathrm{local}}$ to $(i, j')$
        \Else
            \State \textsc{ISend} $\A_{\mathrm{local}}[\,{:}\,r/2,\ :]$ to $(i, j')$
        \EndIf
    \EndFor
\EndIf
\If{Process is in column sender stage ($i = s$)}
    \For{each off-diagonal receiver $(i', j)$ in column $j$}
        \If{Receiver is upper triangle ($i' < j$)}  
            \State \textsc{ISend} $\A_{\mathrm{local}}[\,r/2\,{:},\ :]=(\B_{\mathrm{local}}[:,c/2:])^\top$ to $(i', j)$
        \Else \Comment{lower receiver: full block}
            \State \textsc{ISend} $\A_{\mathrm{local}}=\B_{\mathrm{local}}^\top$ to $(i', j)$
        \EndIf
    \EndFor
\EndIf
\end{algorithmic}}
\end{algorithm}

\subsection{Computation}
\label{sec:impl-comp}

Algorithm~\ref{alg:localcomp} describes the local computation, with $r$ and $c$ denoting the row and column counts of the received blocks.
\new{Halving a CSR matrix along rows and halving it along columns differ in cost, and this asymmetry drives most of the off-diagonal process behavior.} 
Halving along rows is free, since the first half of the row-pointer array already describes a valid submatrix.
Halving along columns is not free, because the first half of the columns is contiguous in neither the column nor the values array, which forces a full traversal and rebuild.
A process computing the top rows of $\C_{ij}$ obtains its input immediately, whereas one computing the right columns must either prune columns directly from the CSR structure or transpose so that columns become rows and pruning is cheap again.

In our experiments, we found that local \TM{MKL} transposes are \TM{at least an order of magnitude faster than MKL SpGEMM kernels}, so the implementation transposes $\B$, builds a handle from the latter half of its row pointers, and then transposes back to obtain the last columns of $\B$.
Because MKL cannot handle the off-by-one inner dimensions that occur when blocks divide unevenly, the inner dimensions of both operands and the output are padded and then trimmed at the end. Diagonal processes ($i = j$) call \texttt{mkl\_sparse\_syrk} directly, while off-diagonal processes call the general routine \texttt{mkl\_sparse\_sp2m}.

\subsection{Communication and Transposes}
\label{sec:impl-transpose}

In Sparse SUMMA, the process broadcasting a block down a column is itself a member of that column, so MPI\_Bcast suffices.
\textsc{CommSpSYRK} instead sends the underlying block of $\A$: at stage $s$, process $(j,s)$ sends $\A_{js}$ down column $j$, playing the role that $\B_{sj} = \A_{js}^\top$ plays in Sparse SUMMA, with the transpose deferred to the local kernel (Algorithm~\ref{alg:commsched}, column-send).
Because the sender is not a member of the receiving column, a collective cannot be used, so individual sends are issued instead.
Given that a sending process may simultaneously be a receiver, MPI\_ISend prevents deadlock.

MKL provides no dedicated transpose routine, so two alternatives were implemented and evaluated: \texttt{mkl\_sparse\_add} with a zero matrix and \texttt{mkl\_sparse\_convert\_csr}. 
The \texttt{mkl\_sparse\_convert\_csr} approach outperformed the \texttt{mkl\_sparse\_add} approach by an order of magnitude, so we adopt it for both the compute-only routine and the full materialization of $\C$. 
If a transpose across the process grid is unavoidable, as in \textsc{SpSYRK}'s input transpose and output materialization, a local transpose is followed by a one-to-one exchange.
\section{Results and Discussion}
\label{sec:results}
This section reports strong scaling and runtime breakdown for each algorithm against appropriate baselines. 
Throughout, ``processes'' refers to MPI processes.

\subsection{Experimental Setup}

\begin{table}[t]
\rev{
    \centering
    \scriptsize
    \setlength{\tabcolsep}{2pt}
    \resizebox{\columnwidth}{!}{%
    \begin{tabular}{@{}lrrrrrrr@{}}
        \toprule
        Label & Rows (M) & Cols (M) & NNZ($\A$) (M) & NNZ($\C$) (M) & Density (\%) & $\gamma$ & Size (GB) \\
        \midrule
        vas\_stokes\_2M   & 2.1  & 2.1  & 65.1  & 1,146.6 & 0.0014   & 18 & 1.5 \\
        stokes            & 11.4 & 11.4 & 349.3 & 5,937.2 & 0.00027  & 17 & 8.4 \\
        kron\_g500-logn21 & 2.1  & 2.1  & 182.1 & 5,772.5 & 0.0041   & 32 & 1.5 \\
        rgg\_n\_2\_24\_s0 & 16.8 & 16.8 & 265.1 & 845.4 & 0.000094 & 3.2 & 2.1 \\
        relat9 & 12.4 & 0.5 & 39.0 & 5,517.0 & 0.00057 & 142 & 0.6\\
        rel9 & 10.0 & 0.3 & 23.7 & 2,038.7 & 0.00087 & 86 & 0.4\\
        spal\_004 & 0.01 & 0.3 & 46.2 & 51.9 & 1.44 & 1.1 & 1.1\\
        \bottomrule
    \end{tabular}}}
    \caption{\new{Test matrices. NNZ($\C$) counts nonzeros in the upper triangle of $\C = \A\A^\T$. Density is $\sfrac{\mathrm{NNZ}(\A)}{mn}$. The output expansion $\gamma = \sfrac{\mathrm{\mathrm{NNZ}(\C)}}{\mathrm{NNZ}(\A)}$ is the ratio of nonzeros in the output to nonzeros in the input.}}
    \label{tab:sizes}
\end{table}

Experiments were run on the CPU nodes of NERSC Perlmutter. 
Each node has 128 CPU cores and 512 GB of memory, connected through a Dragonfly topology. 
Runs used up to 32 nodes with 8 threads \TM{or 4 physical cores} per process and 32 processes per node, varying the number of processes to study strong scaling.
This distribution of processes and threads per node was chosen because we found it to be a sweet spot for per-process compute times, \new{providing the best balance between communication and local computation overhead.}

Two matrix properties govern the cost of sparse SYRK. 
The first is output expansion, $\mathrm{nnz}(\C)/\mathrm{nnz}(\A)$, which determines how much runtime is spent in local kernels and how much memory the result uses. 
The second is the density of the nonzeros, which determines how concentrated the communication is and therefore how much a symmetry-aware schedule can save.
\texttt{vas\_stokes\_2M} and \texttt{stokes} are matrices from \TM{semiconductor process problems} with localized patterns and modest expansion; \texttt{kron\_g500-logn21} is a Kronecker power-law graph whose hub vertices have poor locality and large expansion; and \texttt{rgg\_n\_2\_24\_s0} is a random geometric graph that is spatially localized but has low expansion.
These four cover the two axes independently rather than varying them together.
The remaining three matrices, \texttt{relat9}, \texttt{rel9}, and \texttt{spal\_004}, are rectangular and are used in Section~\ref{sec:runtime-rect} to test the algorithm beyond the square cases; they also span expansion and locality, with \texttt{spal\_004} having low expansion but good locality, and \texttt{relat9} and \texttt{rel9} having similar expansion but different locality.

For the baselines, we used the CombBLAS\TM{} Sparse SUMMA algorithm, except on \texttt{kron} due to out-of-memory errors, and an optimized Sparse SUMMA implementation built on top of CombBLAS that uses MKL kernel. 
In all implementations, matrix values are stored in double precision and the indices in \texttt{int64}.

\subsection{Load Balance}\label{sec:load}

\new{Under load imbalance, \textsc{CommSpSYRK} continued to improve as the number of processes increased, whereas the reference Sparse SUMMA and \textsc{SpSYK} stalled. 
The synchronous broadcasts in Sparse SUMMA force every process to wait for the most heavily loaded one, so even small nonzero imbalances limit scaling, while the point-to-point rounds of \textsc{CommSpSYRK} tolerate imbalance better. 
\textsc{CommSpSYRK}, in its smaller communication rounds, is ``sparsity-aware,'' and therefore improves more under imbalanced conditions; applying a random permutation to the rows and columns of $\A$ restores balance across all approaches.
Therefore, we report load-balanced results wherever possible, except for a rectangular input matrix, where symmetric load balancing is not possible.
\TM{A nonsymmetric permutation is possible but is not offered by CombBLAS, so we do not report those results here.}}
Our results show that load balancing generally improves on standard algorithms while not significantly improving \textsc{CommSpSYRK}, our best algorithm.
As a result, \commspsyrk can skip random permutation as a pre-processing step.

\subsection{Comparison to Baselines}\label{sec:strongscaling}

\begin{figure}
      \centering
      \includegraphics[width=1\linewidth]{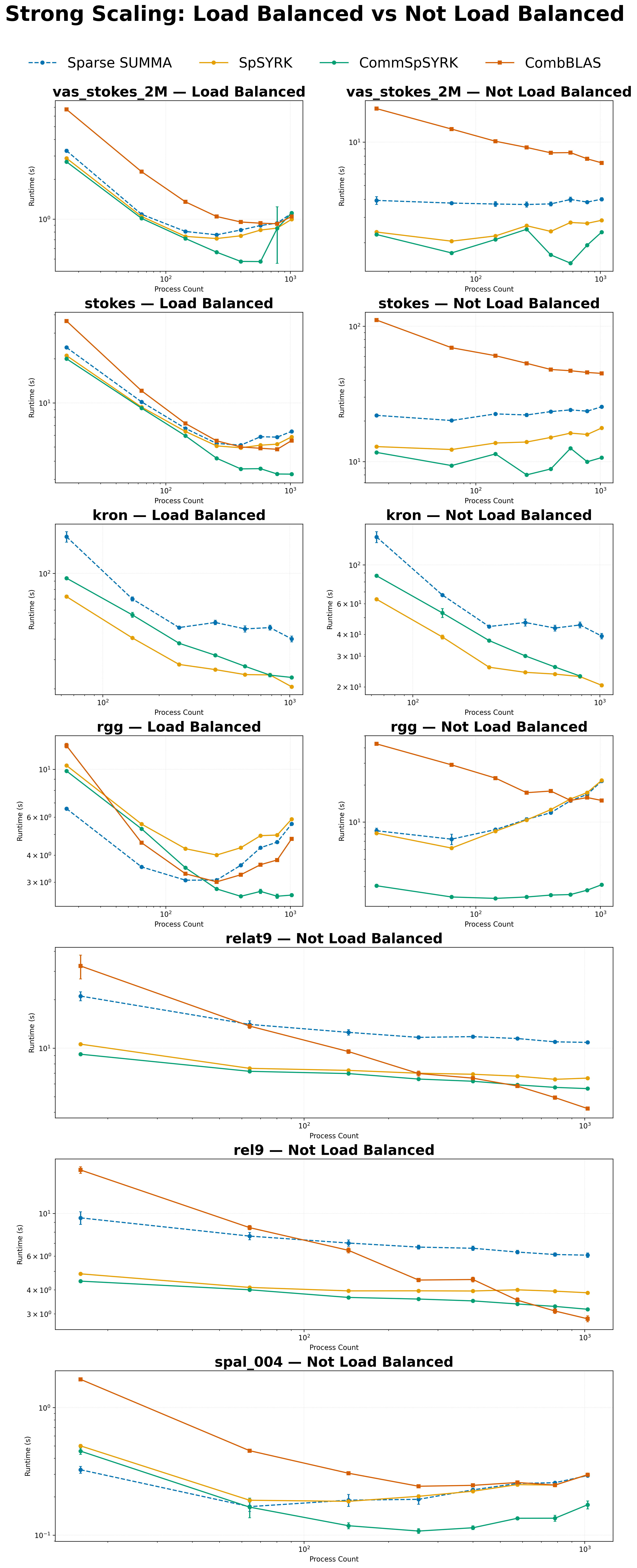}
        \caption{Runtime scaling on each test matrix. Error bars show variation across different runs. CombBLAS runs out of memory on \texttt{kron}, so no data is reported.}
    \label{fig:strong-scaling}
\end{figure}

\new{Figure~\ref{fig:strong-scaling} shows strong-scaling results for all test matrices, with markers indicating mean runtime and error bars representing standard deviation. 
The gap between Sparse SUMMA and \spsyrk varied by matrix: \texttt{rgg} showed parity, \texttt{spal\_004} some regression, and the rest saw roughly a $2\times$ speedup.
Under \textsc{CommSpSYRK} (in comparison to \textsc{SpSYK}), \texttt{spal\_004} and \texttt{rgg} improved significantly, while the rest showed parity or, in the case of \texttt{kron}, a slight regression. This variation follows arithmetic intensity.} 
Halving local work pays off only when local multiplication dominates the runtime, a condition the cost model captures through $\varphi$, the multiply-add count of $\A\A^\top$ (Section~\ref{sec:analysis}).
The matrices with low $\varphi$ relative to $\mathrm{nnz}(\A)$ spend proportionally more time on communication and CSR row and column metadata, neither of which \textsc{SpSYK} changes, so the halved computation is hidden. 
Most matrices scale better with load balancing. 
\texttt{rgg}, which is highly structured (banded), is the clearest example, where a speed regression in most cases becomes a consistent improvement.
In the absence of load balancing, most algorithms show middling performance, except for CombBLAS. 
A likely explanation is that its slower SpGEMM kernel leaves less idle time on its faster processes, so it still achieves a net speedup.

\new{For low process counts, where local computation dominates, \textsc{SpSYK} and \textsc{CommSpSYRK} are comparable and often outperform Sparse SUMMA.
As the process count increases and communication begins to dominate, \textsc{SpSYK} converges toward Sparse SUMMA, matching its communication volume, while \textsc{CommSpSYRK} pulls ahead and approaches the predicted $\sfrac{3}{4}$ ratio.
Our measurements therefore lie in the compute-dominated regime at lower process counts and approach the bandwidth-dominated regime in higher process counts. 
A further breakdown is shown in Section~\ref{sec:runtime-breakdown}.}

\new{For some matrices, notably \texttt{rgg} and \texttt{spal\_004}, the low-process \textsc{SpSYK} gains fall short of the prediction. 
The cause is an inefficiency in the MKL kernel. MKL's local sparse SYRK kernel is less efficient than its SpGEMM kernel on very sparse matrices, so halving the input does not halve the runtime.
Our local benchmarks \TM{on random matrices of varying densities} confirm this, showing that the MKL sparse SYRK kernel is slower than performing the same operation via \texttt{mkl\_sparse\_sp2m} \TM{for matrices with densities less than or equal to $\approx 0.005\%$. 
The slowdown scales with sparsity, with matrices of density lower than $0.0001\%$ consistently showing over $2\times$ slowdown.}}

\subsection{Rectangular Matrices}
\label{sec:runtime-rect}

In our analysis in Section~\ref{sec:analysis}, the row count enters explicitly into $T_{\mathrm{comm}}$ and $T_{\mathrm{tr}}$ through the $m'$ term, whereas the multiply-add count $\varphi$ is determined by the column nonzero distribution rather than by the row count.
Because \textsc{CommSpSYRK} bisects matrices row-wise, the row count decreases at the same rate as the number of nonzeros when they are distributed uniformly across rows; in this regime, the per-process scaling factor for communication and transpose costs is the same for both quantities.
Therefore, a rectangular matrix in our model would exhibit similar performance scaling between Sparse SUMMA and \textsc{CommSpSYRK}. Figure~\ref{fig:strong-scaling} confirms this for \texttt{relat9}, \texttt{rel9}, and \texttt{spal\_004}. At high process counts, CombBLAS outperforms on \texttt{relat9} and \texttt{rel9}, which we suspect arises from the CSC format of CombBLAS giving it a scaling advantage.

\subsection{Communication and Computation Breakdown}
\label{sec:runtime-breakdown}

\begin{figure*}
 \centering
 \includegraphics[width=\linewidth]{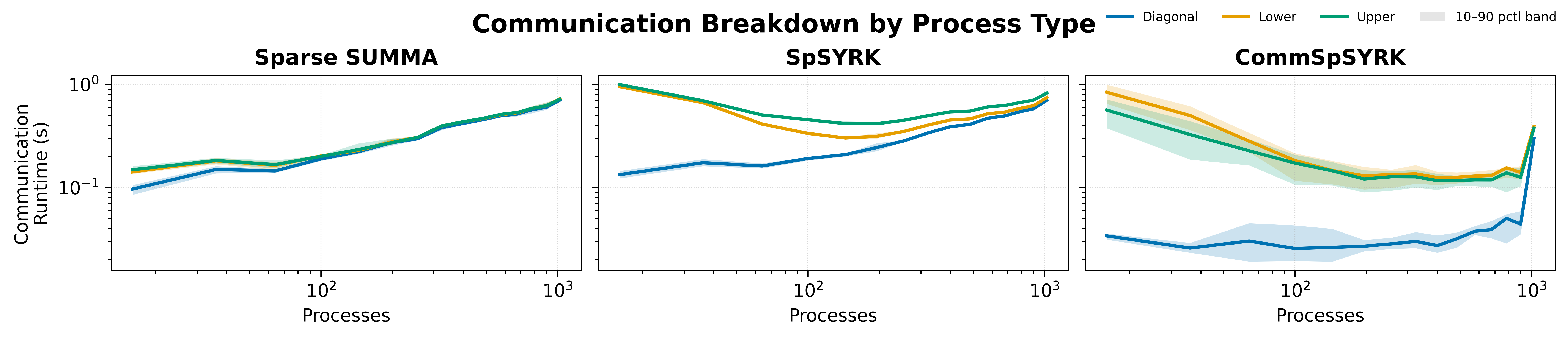}
 \caption{Communication time \rev{on matrix \texttt{rgg} (trends are similar across different matrices)}, broken down by process location. The communication cost of each algorithm is shown, divided by whether a process is on the diagonal of the process grid, the upper triangular, or the lower triangular. \rev{The percentile bands show the distribution over all processes in the runs at that process count.}}
 \label{fig:comm-breakdown}
\end{figure*}

\begin{figure*}
    \centering
           \includegraphics[width=1\linewidth]{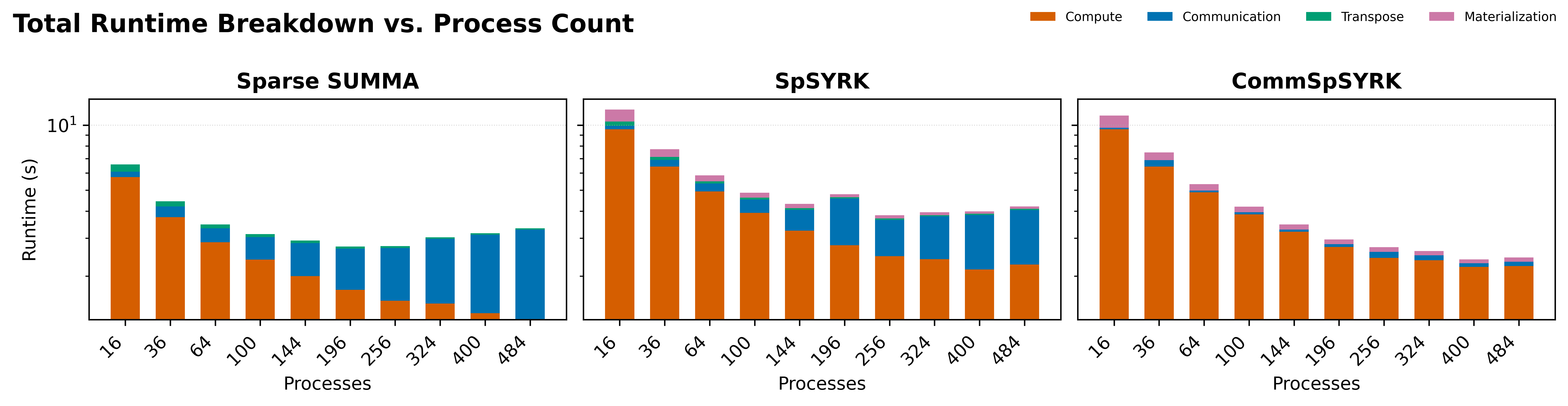}
              \caption{Runtime split by operation for each algorithm. The runtime is divided into computation time, communication time, time spent computing transposes, and time spent materializing the full output. Each bar’s components are determined by the runtime of that component on the processes with the largest total runtime in the grid on \texttt{rgg}.}
    \label{fig:runtime}
\end{figure*}

\new{Figure~\ref{fig:runtime} shows the runtime breakdown: computation in MKL kernels, communication in transfer rounds and staging, and transpose for forming $\A^\top$ and materializing the output.} 
The breakdown is for a single matrix, but the relative trends hold across the entire suite.
Sparse SUMMA and \textsc{SpSYK} show similar trends with increasing process counts and similar scaling, which is expected since Section~\ref{sec:analysis} implies that both achieve constant fractions of the Sparse SUMMA runtime in the high-process limit.

Figure~\ref{fig:comm-breakdown} breaks communication time down by process type: diagonal, upper triangular, and lower triangular, which makes \textsc{CommSpSYRK}'s variance clearer. 
Diagonal processes have lower communication times, receiving only about $\tfrac{2}{3}$ of the data that off-diagonal processes do.
Grouping by type removes much of the variance; the remainder reflects \textsc{CommSpSYRK}'s asynchronous communication, which approaches a synchronous broadcast round as the process count increases.
The stratification between the upper and lower off-diagonals reflects synchronization effects, since the communication system in \textsc{SpSYK} is identical to Sparse SUMMA and the split comes from processes idling while waiting for others to enter the round.
 
\subsection{Memory Footprint and the Materialization Tradeoff}\label{sec:disc-trans}

Figure~\ref{fig:runtime} separates materialization from the rest of the runtime. 
The absolute cost of materialization falls with process count, as Section~\ref{sec:analysis} predicts, since the output transpose scales with the per-process output $\sigma_{\mathrm{mul}}\gamma \eta$. As a fraction of total runtime, it stays roughly stable because the remaining terms shrink at a similar rate.

For the low-process case, the breakdown highlights the relative inefficiency of the MKL sparse SYRK kernel noted in Section~\ref{sec:strongscaling}, since computation there reduces to local multiplication.
For Sparse SUMMA and \textsc{SpSYK}, communication grows until it dominates, and the total runtime increases beyond about 256 processes, whereas \textsc{CommSpSYRK}’s communication stays small and its runtime keeps decreasing over the tested range.

These observations argue against materializing the full output for two reasons.
The first is memory. The output is far denser than the input: $\mathrm{nnz}(\C)$ exceeded $\mathrm{nnz}(\A)$ by $17\times$ to $142\times$ on all matrices except \texttt{rgg} ($3.2\times$) and \texttt{spal\_004} ($1.1\times$) (Table~\ref{tab:sizes}). 
As a result, $\C$, not $\A$, determines the peak memory footprint and thus the largest problem that fits in a given allocation; storing only the triangular half reduces this by half.
The second issue concerns downstream use: consumers are often assumed to need the full matrix, but the motivating applications frequently do not.
For instance, similarity search enumerates each candidate pair once~\cite{guidi2021bella, guidi2021parallel, selvitopi2020distributed}, using only half of the overlap matrix, and the Cholesky factorization of the normal equations reads only a single triangular part. 

In addition, we show that it is possible to perform downstream sparse primitives using only the non-materialized $\C$. Consider the product $\C x$, and let $\C_{\mathrm{L}}$ denote the stored lower-triangular output, including the diagonal, with $D = \mathrm{diag}(\C_{\mathrm{L}})$.
Given that $\C = \C_{\mathrm{L}} + \C_{\mathrm{L}}^{\!\top} - D$,
\begin{equation}
\C x = \C_{\mathrm{L}}\,x + \bigl(x^{\!\top} \C_{\mathrm{L}}\bigr)^{\!\top} - D x,
\label{eq:spmv}
\end{equation}
where the second term uses the transpose of the vector rather than of the matrix, so $\C_{\mathrm{L}}^{\!\top}$ is never formed, and $D x$ removes the double-counted diagonal through a local scaling on diagonal processes.
The arithmetic is unchanged; the only extra cost is that $x$ moves twice, once along process columns and once along process rows, instead of once. 
The same decomposition applies to a matrix operand $X$,
\begin{equation}
\C X = \C_{\mathrm{L}}\,X + \bigl(X^{\!\top} \C_{\mathrm{L}}\bigr)^{\!\top} - D X,
\label{eq:spgemm}
\end{equation}
and here the second term reuses \textsc{CommSpSYRK}'s transposed-send pattern, so the blocks of $\C_{\mathrm{L}}$ it needs are routed by the schedule without additional machinery.

The aggregate data volume of $\C_{\mathrm{L}}$ moved in Eq.~\ref{eq:spgemm} is $2\,\mathrm{nnz}(\C_{\mathrm{L}}) \approx \mathrm{nnz}(\C)$, matching a SUMMA multiply on the materialized matrix, with $X$ still moved once.
Relative to materializing $\C$ and then multiplying, the savings are the output transpose $T_{\mathrm{tr}}$ and half the resident footprint of $\C$, with equal product-phase communication and arithmetic cost.
For $\C x$, the product can instead be computed as $\A(\A^{\!\top}x)$, avoiding $\C$ entirely, so there is no advantage there. 
For $\C X$ the situation is different, since $\A(\A^{\!\top}X)$ can require more arithmetic than $(\A\A^{\!\top})X$, making the triangular form worthwhile. 
Computing the full symmetric completion is thus unnecessary when the output feeds a subsequent kernel, which both lets us evaluate our algorithms without materialization and provides a general method for consuming symmetric output.
\section{Conclusion}

\new{This paper presented \textsc{SpSYRK} and \textsc{CommSpSYRK}, the first distributed-memory algorithms for sparse SYRK that exploit the symmetry of the output matrix. 
\spsyrk divides each off-diagonal block between upper- and lower-triangular process pairs and computes only the lower triangle of each diagonal block, cutting the computation in half compared to Sparse SUMMA.
\textsc{CommSpSYRK} also reorganizes communication so that $\A^\top$ is not formed, reducing communication volume and eliminating the need to transpose the input.
The triangular output can be materialized when a downstream component requires the complete matrix.}

On the Perlmutter supercomputer, the algorithms achieved a 2$\times$ speedup over an optimized Sparse SUMMA on 32 nodes for compute-bound input, with the advantage determined by the arithmetic intensity.
Because the partitioning and communication schemes are agnostic to the local kernel, future work will investigate GPU acceleration and asynchronous execution that overlaps communication and computation.


\section*{Acknowledgments}
This research used resources from the National Energy Research Scientific Computing Center, a DOE Office of Science User Facility supported by the Office of Science of the U.S. Department of Energy under Contract No. DE-AC02-05CH11231, using NERSC award ASCR-ERCAP0030076. 
This material is based upon work supported by the U.S. Department of Energy, Office of Science, Office of Advanced Scientific Computing Research, Department of Energy Computational Science Graduate Fellowship under Award No. DE-SC0025528.
Claude Sonnet 5 was used to generate the code for data collection and the visualizations in Section~\ref{sec:results}.

\bibliographystyle{IEEEtran}
\bibliography{IEEEabrv,ref}

\end{document}